\documentclass[twocolumn,fleqn,aps,pre,showpacs,floatfix,amssymb,superscriptaddress,prb]{revtex4-1}
\usepackage[latin9]{inputenc}
\usepackage{color}
\usepackage{amsmath}
\usepackage{graphicx}
\usepackage[unicode=true,pdfusetitle,
 bookmarks=false,
 breaklinks=false,pdfborder={0 0 1},backref=false,colorlinks=false]
 {hyperref}

\makeatletter
\usepackage[caption=false]{subfig}
\usepackage{color}

\usepackage{graphicx}
\makeatother

\begin{document}
\title{Bridging between Lab and Rotating Frame Master Equations for Open
Quantum Systems }
\author{Gal Shavit}
\affiliation{Raymond and Beverly Sackler School of Physics and Astronomy, Tel Aviv
University, Tel Aviv 6997801, Israel}
\author{Baruch Horovitz}
\affiliation{Department of Physics, Ben-Gurion University of the Negev, Beer Sheva
84105, Israel}
\author{Moshe Goldstein}
\affiliation{Raymond and Beverly Sackler School of Physics and Astronomy, Tel Aviv
University, Tel Aviv 6997801, Israel}
\begin{abstract}
The problem of a driven quantum system coupled to a bath and coherently
driven is usually treated using either of two approaches: Employing
the common secular approximation in the lab frame (as usually done
in the context of atomic physics) or in the rotating frame (prevailing
in, e.g., the treatment of solid-state qubits). These approaches are
applicable in different parts of the parameter space and yield different
results. We show how to bridge between these two approaches by working
in the rotating frame without employing the secular approximation
with respect to the driving amplitude. This allows us to uncover novel
behaviors in regimes which were previously inaccessible or inadequately
treated. New features such as the qualitative different evolution
of the coherence, population inversion at a lower driving amplitude,
and novel structure in the resonance fluorescence spectrum of the
system are found. We argue that this generalized approach is essential
for analyzing hybrid systems, with components that come from distinctly
different regimes which can now be treated simultaneously, giving
specific examples from recent experiments on quantum dots coupled
to optical cavities, and single-spin electron paramagnetic resonance.
\end{abstract}
\maketitle
\section {Introduction}

Complete isolation of any realistic quantum system from its environment
is not typically possible. Open quantum systems, i.e., systems that
have some non-negligible interactions with an external environment,
or ``bath'', evolve over time in a non-unitary fashion, inevitably
leading to processes of relaxation, losses, and phase decoherence
\cite{quantumNoise}. As these dissipative effects can significantly
alter the properties of such systems, the study of open quantum systems
and their evolution over time has important consequences for a myriad
of different quantum devices and applications, perhaps most importantly
in the field of quantum computation and information processing \cite{quantumComputation}.

The dynamics of an open quantum system is typically captured by performing
a perturbative expansion in the bath coupling strength, followed by
integrating out all the bath variables. This leads to a master equation
for the system degrees of freedom, usually represented by the reduced
system density matrix $\rho\equiv\mathrm{Tr}_{B}\left\{ \rho_{SB}\right\} $,
where $\rho_{SB}$ is the total system and bath density matrix, and
$\mathrm{Tr}_{B}\left\{ \text{..}\right\} $ is a trace over the bath
degrees of freedom. One of the most prevalent forms of such equations
is the \textit{Lindblad master equation} \cite{lindbladSemigroups,GoriniGenerators}
\begin{equation}
\frac{d}{dt}\rho=-\frac{i}{\hbar}\left[H_{S},\rho\right]+\mathcal{D}\rho,\label{eq:LindbladMaster}
\end{equation}
where $H_{S}$ its Hamiltonian of the system, controlling the unitary
part of the time evolution, and $\mathcal{D}$ is the dissipator super-operator,
which is of the Lindblad form $\mathcal{D}\rho=\sum_{j}\gamma_{j}\left(L_{j}\rho L_{j}^{\dagger}-\frac{1}{2}\left\{ \rho,L_{j}^{\dagger}L_{j}\right\} \right)$,
with $L_{j}$'s being a set of ``quantum jump'' operators, and $\gamma_{j}$'s
the rates governing the dissipative dynamics. It should be noted that
the Lindblad master equation is the most general form of a master
equation with Markovian dynamics (i.e., without ``memory'' effects)
which preserves the positive semi-definiteness and trace (thus, normalization
of probabilities) of the density matrix \cite{GoriniGenerators,lindbladSemigroups}.
To arrive at \eqref{eq:LindbladMaster} starting from the microscopic
description of the system's dynamics and its interaction with the
bath, one must make the Born-Markov approximation \cite{quantumNoise,breuer2002theory},
which essentially coarse-grains the time evolution of the system density
matrix such that time scales shorter than the bath correlation time
cannot be properly resolved. Another simplification, crucial in obtaining
the Lindbladian form, will be particularly relevant in this work --
\textit{the secular approximation}, requiring that the dissipative
dynamics is sufficiently slow as compared to the system internal time
scale.

The case where the open quantum system is driven out of equilibrium
is of utmost importance in the physics and design of quantum devices.
Such devices must be controlled and manipulated in various ways in
order for them to be useful. Whether preparing a qubit in a specific
desired initial state \cite{RabiOscillationsofExcitonsinSingleQuantumDots},
reading information off it, writing and storing information onto it
\cite{InformationStorageAtom}, probing its current quantum state
\cite{fluxQubitProbe}, or manipulating it by an intricate sequence
of well-designed pulses \cite{DimaPulses,Pulses}, driving of a quantum
system is a hallmark of quantum control and a staple of quantum engineering.

However, the introduction of the driving Hamiltonian into the master
equation is not entirely trivial, as the derivation of the Lindblad
master equation prominently relied on the system Hamiltonian being
diagonal, which is typically not true when a driving term is introduced.
The two conventional and common ways to incorporate the driving effects
can each \textit{potentially lead to different results}, depending
on the parameter regime the system of interest is in. In Sec. \ref{DrivingOpenSystems}
of this paper we review these two different approaches, and discuss
their respective weaknesses and pitfalls. We then introduce a generalized
approach in Sec. \ref{GeneralizedApproach}, which unifies the previous
treatments and extends the range of validity of the quantum master
equation. This more inclusive treatment is similar in essence to the
equations derived in Ref. \cite{PowerFlow}, whose main emphasis was
the thermodynamics of the open system. We explore the consequences
of using the more general approach, when is it superior compared to
the more specialized schemes, what are the appropriate limits in which
our approach converges with this schemes, and importantly, novel qualitative
behavior which may obtained only using the generalized treatment.
These include anomalous time-evolution of the system coherence, population
inversion in unusual regimes, and peculiar structure in the resonance
fluorescence spectrum. The latter, which we study extensively in Sec.
\ref{photolu}, was not previously considered in the context of the
generalized master equation, though it appears to be quite relevant
in distinguishing between different regimes. We also discuss the relevance
of our results to recent experiments on quantum dots coupled to optical
cavities \cite{newMuller,newFlagg,newUlrich,newUlhaqNatPhotonics}
and single electron paramagnetic resonance \cite{STMexperiment,YANG2017,Willke2018}
(a full analysis of the latter type of systems was the subject of
our recent work \cite{ShavitGH}). The latter is accessible in our
approach, contrary to that of Ref. \cite{PowerFlow}, as we further
generalize the nature of the bath itself, and allow it to be out of
thermal equilibrium. Lastly, in Sec. \ref{Caveats} we show when the
proposed treatment may break down, an issue that was not previously
fully examined. We summarize our findings in the Conclusions, Sec.
\ref{Concluz}. Appendix \ref{deriving} details the derivation of
our generalized master equation, while Appendix \ref{triplet} elaborates
on the calculation of two-time correlation functions.

\begin{figure}
\includegraphics[viewport=0bp 0bp 1460bp 1548.604bp,scale=0.19]{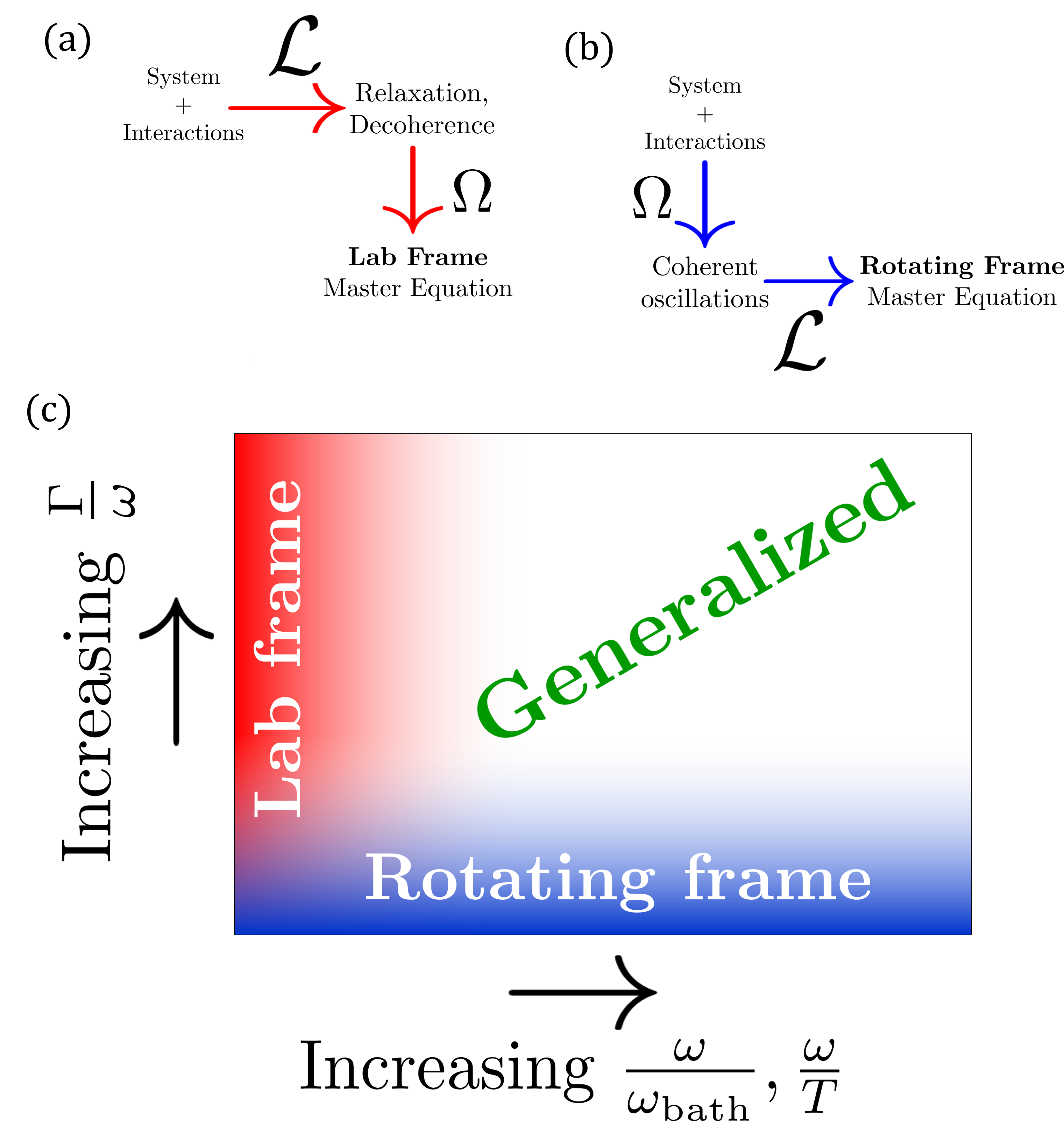}

\caption{\label{fig:non_commute_diagrams} Deriving the master equation for
a driven open quantum system in two different ways. (a) In the lab
frame approach, one first derives the non-driven Lindblad equation,
and then introduces driving to the coherent dynamics. (b) By including
the driving in the system Hamiltonian, and only then deriving the
dissipative time evolution (making a secular approximation with respect
to the driving terms), the rotating frame approach leads to another
distinct form of the master equation. (c) Schematic description of
the range of validity of each of these treatments. Intensity of blue
and red signifies how suitable the lab and rotating frame approaches
are, respectively, in those areas of parameter space. For $\Gamma\ll\omega\ll\omega_{d}$,
the rotating frame approach is adequate, while for $\omega\ll\omega_{{\rm bath}},T\ll\omega_{d}$
(causing only a small modification of the rates), the lab frame is
better suited. Some overlap is apparent, but a region where both approaches
are found wanting can only be be properly addressed by the proposed
generalized approach.}
\end{figure}

\section {Driven open system}\label{DrivingOpenSystems}

For the sake of clarity, but without loss of generality, we will focus
our discussion on two-level systems, or ``qubits''. Prior to adding
the driving term we have the Hamiltonian
\begin{equation}
H=H_{S}^{0}+H_{B}+H_{I},
\end{equation}
with $H_{S}^{0}=-\frac{1}{2}\omega_{0}\sigma_{z}$ (such that the
ground-state has $\sigma_{z}=+1$), $H_{B}$ being the bath Hamiltonian,
and the interaction $H_{I}=\left(a_{x}\sigma_{x}+a_{y}\sigma_{y}+a_{z}\sigma_{z}\right)\hat{B}$,
where the $\sigma$'s are Pauli matrices operating in the qubit Hilbert
space, $\hat{B}$ is some bath operator (assuming different bath operators
couple to each Pauli matrix does not lead to essential modifications
\cite{ShavitGH}) which commutes with the qubit Pauli operators, and
we henceforth set $\hbar=1$. In this simple two-level scenario, the
quantum jump operators in the dissipator are $\sigma_{-}$, $\sigma_{+}$,
and $\sigma_{z}$. The corresponding rates will be denoted by $\gamma_{\downarrow}$,
$\gamma_{\uparrow}$, and $\gamma_{0}$, respectively. These can be
calculated from the bath spectral function as \begin {subequations}
\begin{align}
\gamma_{\downarrow} & \equiv\left(a_{x}^{2}+a_{y}^{2}\right)K\left(\omega_{0}\right),
\end{align}
\begin{equation}
\gamma_{\uparrow}\equiv\left(a_{x}^{2}+a_{y}^{2}\right)K\left(-\omega_{0}\right),
\end{equation}
\begin{equation}
\gamma_{0}\equiv a_{z}^{2}K\left(0\right),
\end{equation}
\end {subequations} with $K\left(\nu\right)\equiv\frac{1}{2}\mathrm{Re}\left\{ \int_{0}^{\infty}d\tau e^{i\nu\tau}\left\langle \hat{B}\left(\tau\right)\hat{B}\left(0\right)\right\rangle _{B}\right\} $
, where $\left\langle \cdot\right\rangle _{B}$ denotes an expectation
value calculated in the bath's stationary state, which is unaffected
by the system in the Born approximation, and will henceforth be assumed
to be a thermal state at temperature $T$ (we take the Boltzmann constant
$k_{B}=1$). The rate calculations are equivalent to Fermi's golden
rule, as the imaginary parts of the bath correlation function contribute
only a Lamb shift to $H_{S}$, not affecting the dissipative dynamics.
Eq. \eqref{eq:LindbladMaster} thus leads to decay and decoherence
of the system, governed by these rates. Note that the secular approximation
would now amount to the assumption that $\omega_{0}$ is much greater
than all $\gamma$'s.

The effects of driving can be best understood by considering continuous
periodic driving, which couples the ground and excited states of the
two-level system. It can be written as an additional time-dependent
term in the system Hamiltonian, $H_{S}=H_{S}^{0}+H_{D}$, with
\begin{equation}
H_{D}=\frac{\Omega}{2}\left(e^{i\omega_{d}t}\sigma_{+}+e^{-i\omega_{d}t}\sigma_{-}\right),\label{eq:drivingHd}
\end{equation}
where $\Omega$ is the driving amplitude, $\omega_{d}\equiv\omega_{0}+\delta\omega$
is the driving frequency ($\delta\omega$ being the detuning), and
we also define the generalized Rabi frequency $\omega\equiv\sqrt{\Omega^{2}+\delta\omega^{2}}$.
We assume a circularly polarized driving field, or else we keep only
its components which appear in Eq. \eqref{eq:drivingHd}, i.e., the
rotating wave approximation (RWA) \cite{rwaLamb}, justified for $\Omega\ll\omega_{0},\omega_{d}$.

As the prescription for deriving Eq. \eqref{eq:LindbladMaster} heavily
relies on the system Hamiltonian being diagonal and time-independent,
the incorporation of such a driving term into the open quantum system
framework is not entirely trivial. There are two common methods of
introducing the driving effects into the Lindbladian master equation,
which mainly differ in when the driving term is taken into account,
as illustrated in Fig. \ref{fig:non_commute_diagrams} and described
in the following.

\subsection {The lab frame approach} 

In this approach the dissipative term in \eqref{eq:LindbladMaster}
is assumed to be the same as with $H_{S}^{0}$, and $H_{D}$ in Eq.
\eqref{eq:drivingHd} is plugged into the commutator term. In other
words, the driving is simply taken as an additional part of the system
Hamiltonian (see Fig. \ref{fig:non_commute_diagrams}a). This procedure,
of inserting the driving Hamiltonian ``after the fact'', i.e., deriving
the dissipative master equation and subsequently changing $H_{S}$,
\textit{is perturbative in $\Omega$}, meaning we would expect it
to be appropriate only when $\Omega$, or more accurately, $\omega$,
is small compared to other relevant energy scales of the system, i.e.,
when
\begin{equation}
\omega\ll\omega_{{\rm bath}},T\ll\omega_{d},\label{eq:labFrameAssumption}
\end{equation}
with the energy scale $\omega_{{\rm bath}}$ defined as the typical
scale over which $K\left(\nu\right)$ remains relatively constant
around $\omega_{d}$, while the temperature $T$ plays a similar role
around $\nu=0$, e.g., in equilibrium detailed balance implies $K\left(\pm\omega\right)=K\left(0\right)\left(1\pm\frac{\omega}{T}\right)$
for $\omega\ll T$. The meaning of $\omega_{{\rm bath}}$ can be illustrated
through an example of a lorentzian spectral density $K\left(\nu\right)\propto\frac{\gamma}{\gamma^{2}+\left(\nu-\nu^{*}\right)^{2}}$,
peaked around some frequency $\nu^{*}$, and having a width $\gamma$,
e.g., a weakly coupled cavity mode \cite{newUlrich}. Expanding $K\left(\omega_{d}\pm\omega\right)$
to first order in $\omega$ around $\omega_{d}$, one finds
\begin{equation}
K\left(\omega_{d}\pm\omega\right)\approx K\left(\omega_{d}\right)\left(1\pm2\frac{\omega}{\omega_{{\rm bath}}}\right),\label{eq:epsEstimate-1}
\end{equation}
where $\omega_{{\rm bath}}=\omega_{d}-\nu^{*}$, and we have assumed
$\gamma\ll\left|\omega_{{\rm bath}}\right|$.

Representing the system reduced density matrix as 
\begin{equation}
\rho\equiv\begin{pmatrix}1-n & \alpha\\
\alpha^{*} & n
\end{pmatrix}=\frac{1}{2}+\frac{1-2n}{2}\sigma_{z}+\alpha^{*}\sigma_{-}+\alpha\sigma_{+},\label{eq:rhoSdef}
\end{equation}
the master equation is fully described by (in an interaction frame
rotating with frequency $\omega_{d}$) \begin {subequations}
\begin{equation}
\frac{d}{dt}n=-\left(\gamma_{\downarrow}+\gamma_{\uparrow}\right)n+\gamma_{\uparrow}-i\Omega\frac{\alpha-\alpha^{*}}{2},\label{eq:labframeN}
\end{equation}
\begin{equation}
\frac{d}{dt}\alpha=-\left(\tilde{\gamma}+i\delta\omega\right)\alpha-i\Omega\left(n-\frac{1}{2}\right),\label{eq:labframeAlpha}
\end{equation}
\end {subequations} with $\tilde{\gamma}\equiv\frac{\gamma_{\downarrow}+\gamma_{\uparrow}}{2}+2\gamma_{0}$.
The more commonly known decay times characterizing the evolution of
the qubit are then $\frac{1}{T_{1}}=\gamma_{\downarrow}+\gamma_{\uparrow},$
$\frac{1}{T_{2}^{*}}=2\gamma_{0}$, and $\frac{1}{T_{2}}=\frac{1}{2T_{1}}+\frac{1}{T_{2}^{*}}$.
This ``lab frame'' approach is popular mainly in atomic physics \cite{MollowRef4,AtomicRhobidium}
and quantum optics \cite{QuantumOptics}, where the amplitude of the
driving field may indeed be several orders of magnitude weaker compared
to the qubit energy scale. This result is better known as the famous
Bloch equations \cite{BlocEqns}, which are entirely equivalent to
\eqref{eq:labframeN}--\eqref{eq:labframeAlpha} using the proper
definitions.

\subsection {The rotating frame approach} 

Unlike the lab frame approach, the rotating frame approach does take
into account the effects of stronger driving on the dissipator. Here,
the full system Hamiltonian is diagonalized first, and only then the
Lindblad formalism takes place in the usual way (Fig. \ref{fig:non_commute_diagrams}b).
This approach is most common in solid-state systems, e.g., superconducting
qubits \cite{ShnirmanRotating,RotatingCooperPair,RotatingQubit},
where the driving cannot be dealt with in a perturbative manner. 

Upon applying a transformation to a rotating frame by defining the
unitary operator $U\equiv e^{-\frac{i\omega_{d}t}{2}\sigma_{z}}$
and transforming $H\rightarrow UHU^{\dagger}-iU\dot{U}^{\dagger}$,
so as to eliminate the time dependence in $H_{D}$, and then diagonalizing
the system Hamiltonian in this rotating frame, one arrives at the
transformed Hamiltonian
\begin{equation}
\tilde{H}=\frac{1}{2}\omega\sigma_{z}+\tilde{H}_{I}+H_{B}.\label{eq:rotatingFrameHamiltonian}
\end{equation}
A key factor here is the modification the system-bath interaction
Hamiltonian by the rotation $U$ and by diagonalizing the system Hamiltonian.
In the frame of Eq. \eqref{eq:rotatingFrameHamiltonian}, defining
$\tilde{\rho}\equiv\begin{pmatrix}d & x\\
x^{*} & u
\end{pmatrix}$ with $d=1-u$ yields the Lindblad master equation\begin {subequations}
\begin{equation}
\frac{d}{dt}u=-\left(\kappa_{\uparrow}+\kappa_{\downarrow}\right)u+\kappa_{\uparrow},\label{eq:rotMaster1}
\end{equation}
\begin{equation}
\frac{d}{dt}x=-\left(\frac{\kappa_{\uparrow}+\kappa_{\downarrow}}{2}+\kappa^{*}+i\omega\right)x,\label{eq:rotMaster2}
\end{equation}
\end {subequations} with rates defined as \begin {subequations}
\begin{align}
\kappa_{\uparrow} & =\frac{\left(\sin\beta-1\right)}{4}^{2}\Gamma_{\downarrow}\left(1-\epsilon\right)+\frac{\left(\sin\beta+1\right)}{4}^{2}\Gamma_{\uparrow}\left(1+\epsilon_{e}\right)\nonumber \\
 & +\cos^{2}\beta\Gamma_{-}^{z},
\end{align}
\begin{align}
\kappa_{\downarrow} & =\frac{\left(\sin\beta+1\right)}{4}^{2}\Gamma_{\downarrow}\left(1+\epsilon\right)+\frac{\left(\sin\beta-1\right)}{4}^{2}\Gamma_{\uparrow}\left(1-\epsilon_{e}\right)\nonumber \\
 & +\cos^{2}\beta\Gamma_{+}^{z},
\end{align}

\begin{equation}
\kappa^{*}=2\sin^{2}\beta\Gamma_{0}^{z}+\frac{\cos^{2}\beta}{2}\left(\Gamma_{\downarrow}+\Gamma_{\uparrow}\right),
\end{equation}
\end {subequations} where $\tan\beta\equiv\frac{\delta\omega}{\Omega}$,
$\Gamma_{\downarrow}\equiv\left(a_{x}^{2}+a_{y}^{2}\right)K\left(\omega_{d}\right)$,
$\Gamma_{\uparrow}\equiv\left(a_{x}^{2}+a_{y}^{2}\right)K\left(-\omega_{d}\right)$,
$\Gamma_{0}^{z}=\gamma_{0}$, $\Gamma_{\pm}^{z}\equiv a_{z}^{2}K\left(\pm\omega\right)$,
and $\epsilon,\epsilon_{e}$ are defined by a first-order in $\omega$
expansion of the spectral density, $K\left(\omega_{d}\pm\omega\right)=K\left(\omega_{d}\right)\left[1\pm\epsilon\right]$,
and $K\left(-\omega_{d}\pm\omega\right)=K\left(-\omega_{d}\right)\left[1\pm\epsilon_{e}\right]$.
One typically expects $\epsilon,\epsilon_{e}\sim\omega/\omega_{{\rm bath}}$.
This approach evidently gives rise to a much richer landscape of rates
governing the dissipation, with more bath spectral components appearing
in the dissipator. However, an important key distinction of this approach
compared to the lab frame is that in the transition from Eq. \eqref{eq:rotatingFrameHamiltonian}
to \eqref{eq:rotMaster1}--\eqref{eq:rotMaster2}, we used a \textit{secular
approximation with regards to $\omega$}, i.e., demanding
\begin{equation}
\Gamma_{\uparrow/\downarrow},\Gamma_{0/\pm}^{z}\ll\omega\ll\omega_{d}.\label{eq:RotFramAssumptions}
\end{equation}

In deciding which of the presented approaches to use when interested
in the time evolution of a driven open quantum system, one must examine
the validity of \eqref{eq:labFrameAssumption} or \eqref{eq:RotFramAssumptions},
see Fig. \ref{fig:non_commute_diagrams}c. Although in some cases
at least one of the assumptions may be appropriate, there are scenarios
in which both are inadequate and will inevitably compromise the validity
of the calculated dynamics. 

\section {A generalized approach}\label{GeneralizedApproach}

A more inclusive \textit{generalized approach}, where the aforementioned
extra assumptions are not necessary, may be obtained by following
the rotating frame method, and omitting the final secular approximation.
Thus, the only assumption is that the bare qubit frequency $\omega_{0}$
and the drive frequency $\omega_{d}$ are much larger than all other
energy scales. This introduces additional non-secular terms into the
master equation for $\frac{d}{dt}\tilde{\rho}$, which in the rotating
frame are time-dependent with oscillatory factors going as $e^{\pm i\omega t}$
and $e^{\pm2i\omega t}$ (see Appendix \ref{deriving}). Then, by
performing a unitary transformation $\rho\to e^{-i\tilde{H}_{S}t}\rho e^{i\tilde{H}_{S}t}$
(where $\tilde{H}_{S}=\frac{1}{2}\omega\sigma_{z}$), we eliminate
these undesirable time dependencies. We find
\begin{eqnarray}
\frac{d}{dt}n & = & -n\left(\Gamma_{\downarrow}\left(1+\epsilon\sin\beta\right)+\Gamma_{\uparrow}\left(1+\epsilon_{e}\sin\beta\right)\right)\nonumber \\
 &  & +\Gamma_{\uparrow}\left(1+\epsilon_{e}\sin\beta\right)-i\Omega\frac{\alpha-\alpha^{*}}{2}\nonumber \\
 &  & -\frac{\alpha+\alpha^{*}}{2}\cos\beta\frac{\Gamma_{\downarrow}\epsilon+\Gamma_{\uparrow}\epsilon_{e}}{2},\label{eq:generalN}
\end{eqnarray}
\begin{eqnarray}
\frac{d}{dt}\alpha & = & -\alpha\left(\tilde{\Gamma}+i\delta\omega\right)-i\Omega\left(n-\frac{1}{2}\right)\nonumber \\
 &  & +\left(n-\frac{1}{2}\right)\sin\beta\cos\beta\left(\Gamma_{+}^{z}+\Gamma_{-}^{z}-2\Gamma_{0}^{z}\right)\nonumber \\
 &  & -\frac{\Gamma_{\downarrow}\epsilon-\Gamma_{\uparrow}\epsilon_{e}}{4}\cos\beta-\frac{\Gamma_{+}^{z}-\Gamma_{-}^{z}}{2}\cos\beta,\label{eq:generalA}
\end{eqnarray}
with $\tilde{\Gamma}\equiv\frac{\Gamma_{\downarrow}}{2}\left(1+\epsilon\sin\beta\right)+\frac{\Gamma_{\uparrow}}{2}\left(1+\epsilon_{e}\sin\beta\right)+\left(\Gamma_{+}^{z}+\Gamma_{-}^{z}\right)\cos^{2}\beta+2\Gamma_{0}^{z}\sin^{2}\beta$.
Eqs. \eqref{eq:generalN}--\eqref{eq:generalA} are the main result
of this paper. One immediately recognizes that taking the approximations
$K\left(\pm\omega_{d}\pm\omega\right)\rightarrow K\left(\pm\omega_{0}\right)$
and $K\left(\pm\omega\right)\rightarrow K\left(0\right)$, leading
to $\epsilon=\epsilon_{e}=0$ and $\Gamma_{+}^{z}=\Gamma_{-}^{z}=\Gamma_{0}^{z}$,
recovers the lab frame Eqs. \eqref{eq:labframeN}--\eqref{eq:labframeAlpha}.
This approximation indeed requires the generalized Rabi frequency
to be small enough, in the sense of Eq. \eqref{eq:labFrameAssumption}. 

\begin{figure}
\begin{centering}
\includegraphics[scale=0.6]{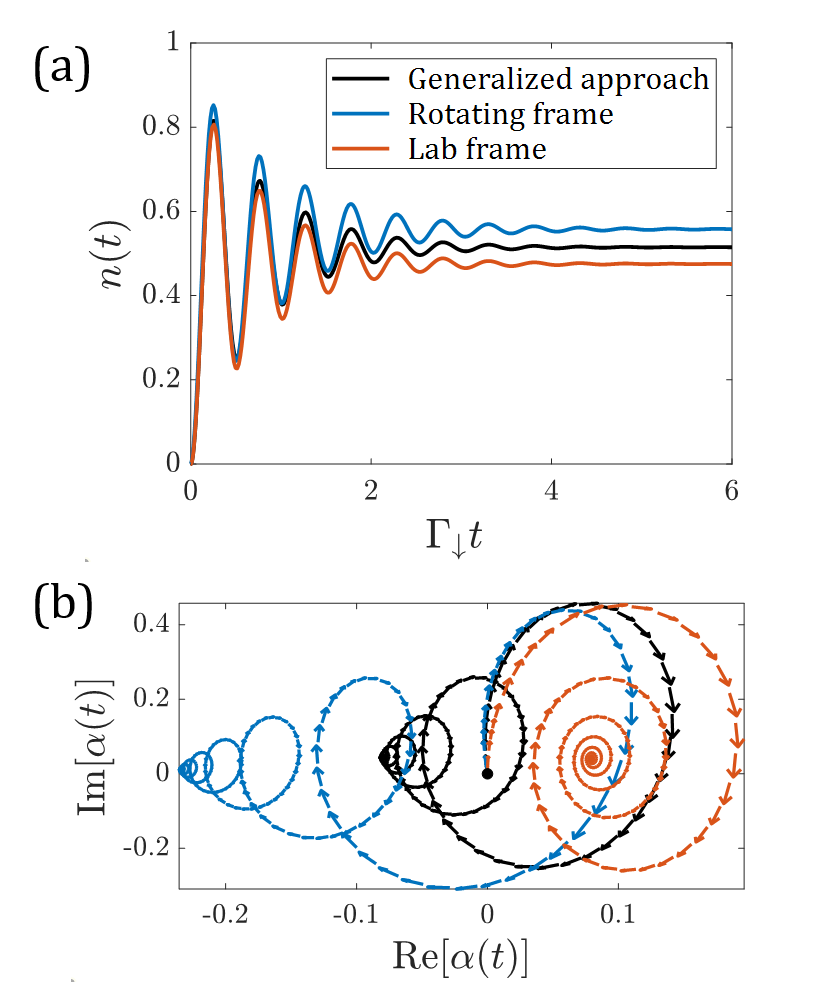}
\par\end{centering}
\caption{\label{fig:transientim} Time evolution of the density matrix components
under the three different approaches discussed: lab frame (red), rotating
frame (blue), and the generalized approach (black). (a) The time evolution
of the excited state population. (b) Evolution of the off-diagonal
component $\alpha$ in the complex plane, beginning at $\left(0,0\right)$
(large black dot) and following the direction of the arrows. Initially
the qubit is in the pure two-level ground state ($n\left(0\right)=0$,
$\alpha\left(0\right)=0$), and the parameters used are $\Omega=12\Gamma_{\downarrow}$,
$\delta\omega=3\Gamma_{\downarrow}$, $\Gamma_{0}^{z}=\frac{1}{2}\Gamma_{\downarrow}$,
$\Gamma_{\pm}^{z}=\Gamma_{0}^{z}\left(1\pm\frac{1}{2}\right)$, $\epsilon=0.1$,
and $\Gamma_{\uparrow}=\epsilon_{e}=0$.}
\end{figure}

As a first example of the difference in the resulting open system
dynamics when employing the various approaches, the time evolution
of the driven qubit is numerically calculated and shown in Fig. \ref{fig:transientim}
for each approach. As the calculations take place in a regime where
neither \eqref{eq:labFrameAssumption} nor \eqref{eq:RotFramAssumptions}
are adequate (the generalized Rabi frequency is comparable to the
decay rates, and we introduce finite asymmetry $\epsilon$ to the
spectral components that determine the rates), we expect the qubit
``trajectory'' in the previous Lindbladian approaches to deviate from
our less restrictive general calculation. While the differences in
$n\left(t\right)$, the excited state population, are moderate, the
departure from the generalized result in the evolution of the coherence
$\alpha$, for both the lab and rotating frame approaches, is much
more dramatic, both in terms of the steady state value and the evolution
towards it. This component of the density matrix is commonly the one
of most interest for quantum devices.

\begin{figure}
\begin{centering}
\includegraphics[viewport=33.5941bp 0bp 454bp 451bp,scale=0.6]{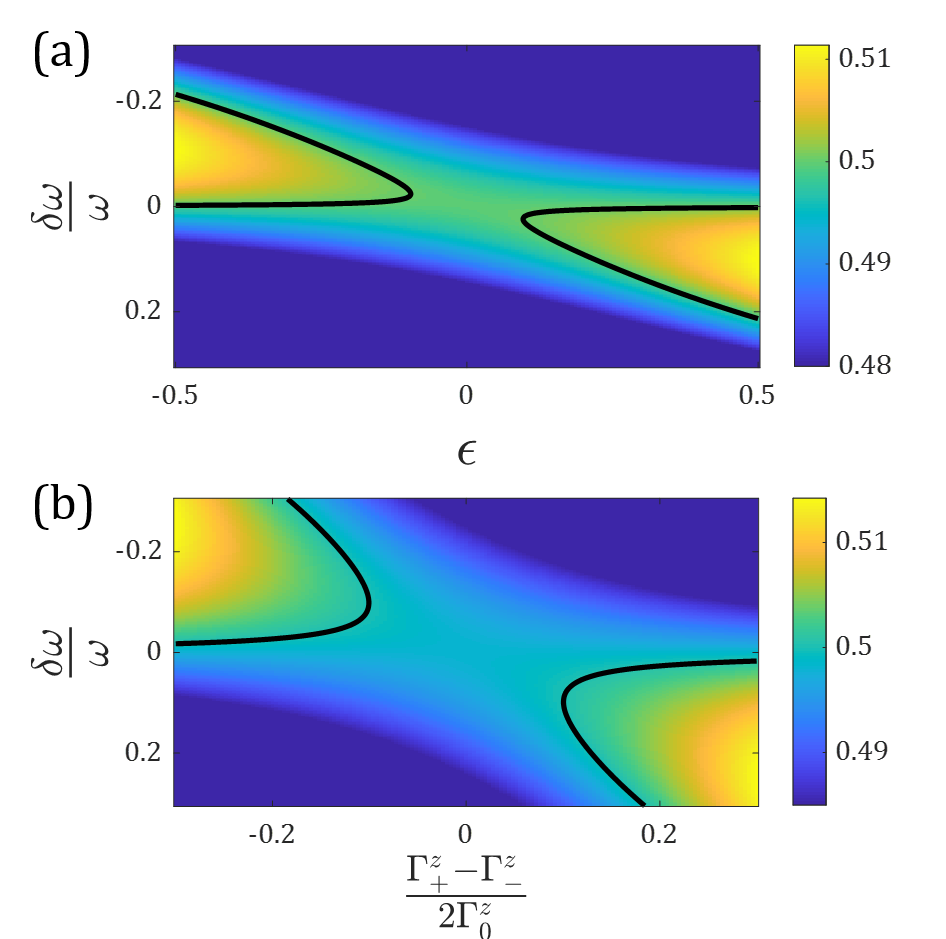}
\par\end{centering}
\caption{\label{fig:PopInverse} Steady-state excited state qubit population
$n$ in the presence of spectral asymmetry of the bath correlation
function $K$. (a) $n$ as a function of $\epsilon$ and detuning,
with $\Gamma_{0}^{z}=\frac{1}{20}\Gamma_{\downarrow}$ and $\Gamma_{+}^{z}=\Gamma_{-}^{z}=\Gamma_{0}^{z}$.
(b) $n$ as a function of the longitudinal asymmetry $\frac{\Gamma_{+}^{z}-\Gamma_{-}^{z}}{2\Gamma_{0}^{z}}$
and detuning, with $\epsilon=0$, $\Gamma_{0}^{z}=\Gamma_{\downarrow}$,
and $\Gamma_{+}^{z}+\Gamma_{-}^{z}=2\Gamma_{0}^{z}$. The areas within
the solid black lines are where population inversion occurs, $n>\frac{1}{2}$,
and the color scale is chosen as to accentuate the effect. Other parameters
used: $\Gamma_{\uparrow}=\epsilon_{e}=0$, $\Omega=25\Gamma_{\downarrow}$.}
\end{figure}

\begin{figure}
\begin{centering}
\includegraphics[viewport=28.81423bp 0bp 486bp 799bp,scale=0.6]{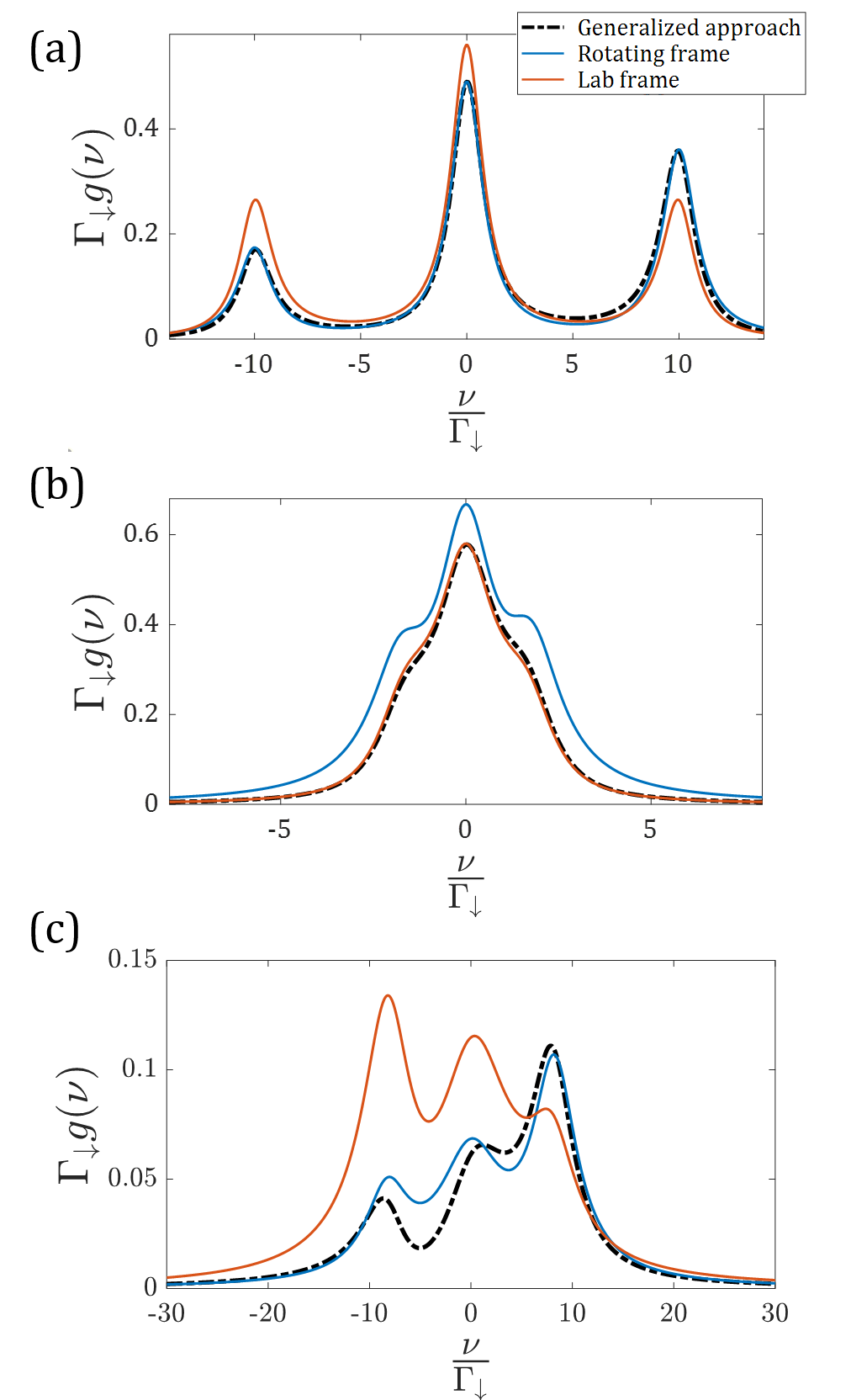}
\par\end{centering}
\caption{\label{fig:tripletim} $g\left(\nu\right)$ calculated using the different
approaches discussed: lab frame (red), rotating frame (blue) and generalized
approach (black) for different parameter sets: (a) $\Omega=1.8\Gamma_{\downarrow}$,
$\Gamma_{0}^{z}=0.2\Gamma_{\downarrow}$, $\Gamma_{\pm}^{z}=\Gamma_{0}^{z}\left(1\pm\frac{\omega}{T}\right)$,$T=12.5\Gamma_{\downarrow}$,
and $\delta\omega=\Gamma_{\uparrow}=\epsilon_{e}=\epsilon=0$. (b)
Same as (a), except for $\Omega=10\Gamma_{\downarrow}$. (c) $\Omega=8\Gamma_{\downarrow}$,
$\delta\omega=-2\Gamma_{\downarrow}$, $\Gamma_{0}^{z}=2\Gamma_{\downarrow},$
$\Gamma_{+}^{z}=3\Gamma_{\downarrow},$ $\Gamma_{0}^{z}=0.4\Gamma_{\downarrow},$and
$\Gamma_{\uparrow}=\epsilon_{e}=\epsilon=0$.}
\end{figure}

The importance of the spectral asymmetry of $K$ around the driving
frequency $\omega_{d}$ and the dc component already becomes clear
at the steady-state level for $n$, namely the possibility of \textit{population
inversion in the qubit}, even in the regime $\Gamma_{\uparrow}<\Gamma_{\downarrow}$,
see Fig. \ref{fig:PopInverse}. While this effect is possible in the
rotating frame approach, it is completely absent from the lab frame
one. The generalized treatment allows us to explore such a remarkable
effect without requiring any unnecessary assumptions on the size of
the decay rates relative to $\omega$. Thus, even in systems where
Eq. \eqref{eq:RotFramAssumptions} is somewhat inadequate, population
inversion may be detected as a possible signature of spectral asymmetry.

\section {Resonance fluorescence}\label{photolu}

An interesting quantity often measured in experiments is the correlation
function
\begin{equation}
g\left(\nu\right)\equiv\int_{-\infty}^{\infty}d\tau e^{i\nu\tau}\left\langle \sigma_{-}\left(t_{0}\right)\sigma_{+}\left(t_{0}+\tau\right)\right\rangle ,
\end{equation}
representing resonance fluorescence \cite{corrGlauber,MollowTriplet},
i.e., the cross section for inelastic scattering of photons near resonance.
It typically features the Mollow triplet of peaks at the original
frequency $\omega_{d}$ and at the dressed frequencies $\omega_{d}\pm\omega$.
By utilizing the quantum regression theorem \cite{QrEGRESSION}, we
perform calculations with the three different approaches (see Appendix
\ref{triplet}) in two distinct parameter regimes, distinguished by
the size of the driving amplitude $\Omega$. In the first regime,
$\Omega$ is sufficiently large such that it causes a significant
shift in $\Gamma_{\pm}^{z}$ away from $\Gamma_{0}^{z}$, making \eqref{eq:labFrameAssumption}
inadequate. This results in the asymmetry of the triplet shown in
Fig. \ref{fig:tripletim}a. In this large $\Omega$ regime the rotating
frame approach approximates well the generalized one, while the lab
frame is not reliable as all the effects of the driving on the dissipator
are neglected. In the other regime, $\Omega$ is smaller and comparable
in size to the decay rates, such that \eqref{eq:RotFramAssumptions}
is violated. In that case the rotating frame approach is not applicable
since the additional secular approximation is not well justified.
This discrepancy is clearly visible in Fig. \ref{fig:tripletim}b,
where the lab frame result matches fairly well the generalized one.

The generalized approach, as explained before, allows for a more complete
description of the dynamics in both of these extreme regimes, and
in all the parameter range in between. The calculation presented in
Fig. \ref{fig:tripletim}c is an example of such an intermediate regime.
In this scenario the different parameters are chosen such that $\omega$
is comparable with both the decay rates \textit{and} the energy scales
determining the spectral asymmetry, $\omega_{\mathrm{bath}}$ and
$T$. In Fig. \ref{fig:tripletim}c the complex stsructure of the
two side peaks is considerably misrepresented by either the lab or
rotating-frame treatments. The asymmetric generalized result has significabtly
greater range of applicability in the chosen parameter regime, and
is able to capture the interplay between the detuning, which ``favors''
lower frequencies since it is negative, and the spectral asymmetry,
having the opposite effect due to $\Gamma_{+}^{z}>\Gamma_{0}^{z}$.

\begin{figure*}
\begin{centering}
\includegraphics[viewport=72.0024bp 0bp 1189bp 339bp,scale=0.5]{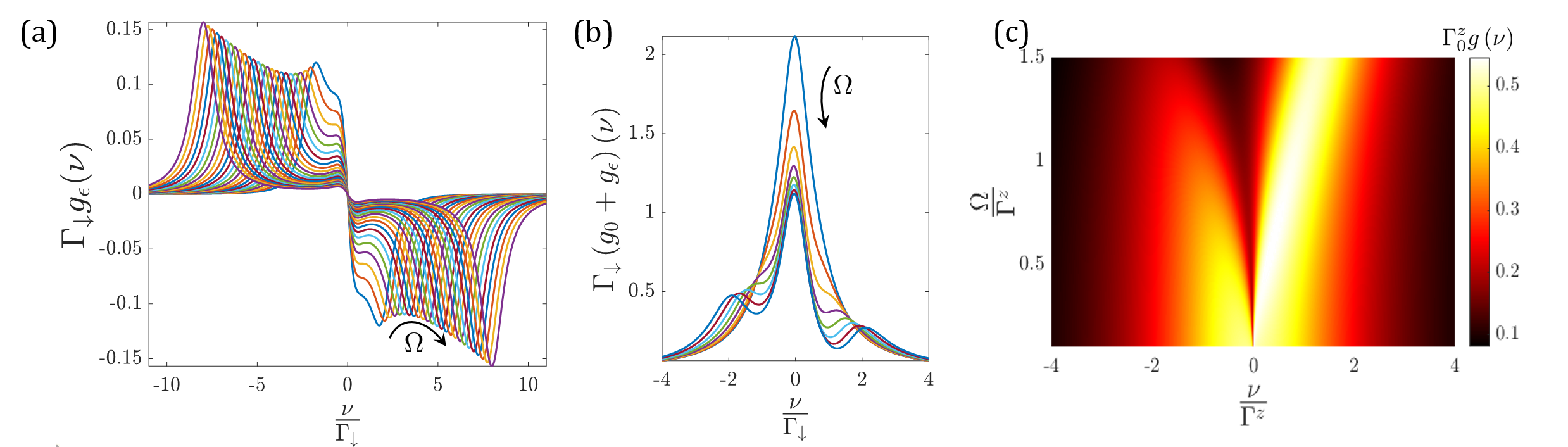}
\par\end{centering}
\caption{\label{fig:InterestingRegimes} Relevant experimental signatures in
regimes where only the generalized approach is free
from inadequate approximations on the system dynamics. (a) The asymmetric
correction to the resonance fluorescence spectrum $g_{\epsilon}\left(\nu\right)$,
calculated on resonance for zero longitudinal coupling. The values
of $\frac{\Omega}{\Gamma_{\downarrow}}$ we use vary from $2$ to
$8$, in steps of $0.25$ (the direction of increasing $\Omega$ is
indicated by an arrow). (b) The full corrected lineshape in the regime
$\Omega\sim\Gamma_{\downarrow}$. $\frac{\Omega}{\Gamma_{\downarrow}}$
varies from $0.5$ to $2$ with steps of 0.2 (the direction of increasing
$\Omega$ is indicated by an arrow). In (a),(b) we use $\epsilon\approx\frac{\omega}{\omega_{{\rm bath}}}$,
with $\omega_{{\rm bath}}=-20\Gamma_{\downarrow}$. (c) $\Gamma_{0}^{z}g\left(\nu\right)$
in the large dephasing regime, $\frac{1}{T_{1}}\ll\Omega\lesssim\frac{1}{T_{2}}$,
on resonance, with different driving amplitudes. Here we use the asymmetry
parameter $\Delta=0.4$.}
\end{figure*}

Interestingly, the resonance fluorescence spectra of semiconductor
quantum dots in optical cavities, which comprise effective two-level
systems, have been extensively measured \cite{newMuller,newFlagg,newUlrich,newUlhaqNatPhotonics,newUlhaqOpEx}.
Our generalized approach is particularly relevant to such experiments
since (i) the driving amplitudes used are typically not much greater
than the dissipative energy scale which determines the linewidths,
such that the rotating frame approach is somewhat inadequate, and
(ii) the spectral density of the qubit bath is heavily influenced
by the proximity of a cavity mode to the quantum dot resonance. This
is especially important in the limit of weak coupling to a cavity,
where the loss rate of the cavity is much higher than the coupling
constant to the quantum dot, see, e.g., Ref. \cite{newUlrich}. In
that case, instead of using the full Jaynes-Cummings model to account
for the cavity mode, we can incorporate it as a bath with transverse
coupling and a peaked spectral density, arriving at the situation
described by Eq. \eqref{eq:epsEstimate-1}, i.e., $\epsilon\approx\frac{\omega}{\omega_{{\rm bath}}}$.
This can make the spectral functions $K\left(\pm\omega_{d}-\omega\right)$,
$K\left(\pm\omega_{d}\right)$, and $K\left(\pm\omega_{d}+\omega\right)$
significantly different from one another even when $\omega\ll\omega_{d}$,
and lead to an apparent asymmetry in the resonance fluorescence spectrum,
similarly to Fig. \ref{fig:tripletim}. For the simplified case of
zero longitudinal coupling and $\Gamma_{\uparrow}=0$, the resonance
fluorescence spectrum on resonance is given by $g=g_{0}\left(\nu\right)+g_{\epsilon}\left(\nu\right)+{\cal O}\left(\epsilon^{2}\right)$,
with $g_{0}\left(\nu\right)$ being the lab frame result, and $\nu$
is the scattered frequency in the rotating frame (i.e. it is $\nu+\omega_{d}$
in the lab frame). The result on resonance $\delta\omega=0$ is \begin{widetext}
\begin{equation}
g_{0}\left(\nu\right)=4\Gamma_{\downarrow}\frac{\Gamma_{\downarrow}^{2}\left(\Gamma_{\downarrow}^{2}+\nu^{2}\right)\left(\Gamma_{\downarrow}^{2}+4\nu^{2}\right)+2\Omega^{2}\left(\left(\Gamma_{\downarrow}^{2}+\Omega^{2}\right)^{2}-\left(\Gamma_{\downarrow}^{2}+2\Omega^{2}\right)\nu^{2}+4\nu^{4}\right)}{\left(\Gamma_{\downarrow}^{2}+4\nu^{2}\right)\left(\Gamma_{\downarrow}^{2}+2\Omega^{2}\right)\left[\Gamma_{\downarrow}^{2}\left(\Gamma_{\downarrow}^{2}+4\Omega^{2}+5\nu^{2}\right)+4\left(\nu^{2}-\Omega^{2}\right)^{2}\right]},
\end{equation}
\begin{equation}
g_{\epsilon}\left(\nu\right)=2\epsilon\Gamma_{\downarrow}\Omega\nu\frac{\Gamma_{\downarrow}^{2}\left(13\Gamma_{\downarrow}^{2}+11\Omega^{2}\right)+4\nu^{2}\left(\Gamma_{\downarrow}^{2}+3\Omega^{2}\right)}{\left(\Gamma_{\downarrow}^{2}+4\nu^{2}\right)\left(\Gamma_{\downarrow}^{2}+2\Omega^{2}\right)\left[\Gamma_{\downarrow}^{2}\left(\Gamma_{\downarrow}^{2}+4\Omega^{2}+5\nu^{2}\right)+4\left(\nu^{2}-\Omega^{2}\right)^{2}\right]}.\label{eq:gepsilon}
\end{equation}
\end {widetext} The correction to the resonance fluorescence spectrum
is valid for arbitrary $\frac{\Omega}{\Gamma_{\downarrow}}$ within
the generalized treatment, and thus could not be reliably obtained
by the usual rotating frame treatment. Notice that $g_{\epsilon}\left(\nu\right)$
will lead to an asymmetry, due to it being an odd function of $\nu$,
see Fig. \ref{fig:InterestingRegimes}a. For $\Omega\gg\Gamma_{\downarrow}$,
$g_{\epsilon}$ is peaked mainly around $\pm\Omega$, yet a crossover
into a regime where only our generalized approach is valid is clearly
apparent when $\Omega$ is decreased. For this smaller driving amplitude
regime, the asymmetry may begin to be visibly pronounced in the central
peak as well, as can be seen in Fig. \ref{fig:InterestingRegimes}b.
The lower driving amplitude regime shows some resemblance to the results
presented in Ref. \cite{newUlrich} (lower curves in Fig. 2a).

This asymmetric effect is in line with the measurements made in Refs.
\cite{newFlagg,newUlrich,newUlhaqNatPhotonics}, which showed an asymmetry
which grows with the driving amplitude. In Ref. \cite{newUlrich},
a triplet asymmetry of $15\%$ was reported, consistent with the proximity
to the cavity mode $\frac{\omega}{\omega_{{\rm bath}}}\sim0.1-0.3$.
A small amount of asymmetry could also be noticed in the central triplet
peak in some of the plots of Ref. \cite{newUlrich} with smaller $\Omega$,
in agreement with our theoretical predictions. The possible effect
of the cavity was actually pointed out in Ref. \cite{newRoyHughs}.
However, that work utilizes an analog of the rotating frame approach,
\textit{which is not entirely justified} in this particular experimental
regime, though it provides a decent qualitative description. In Ref.
\cite{newMuller} the Rabi frequency is even tuned all the way down
below $T_{2}^{-1}$, prompting its Authors to employ the restrictive
lab frame approach. Thus, we conclude that the generalized approach
is needed to account for the behavior of common cavity-coupled quantum
dots, as it is the only one capturing the asymmetric features without
imposing inadequate approximations.

We note that the so-called excitation-induced dephasing (EID) effect
may also be captured using our generalized approach. By expanding,
e.g., $K\left(\omega_{d}\pm\omega\right)$ to second order in $\frac{\omega}{\omega_{{\rm bath}}},$
additional terms are added to our master equation \eqref{eq:generalN}--\eqref{eq:generalA},
among them is the change of the dephasing rate $\tilde{\Gamma}\rightarrow\tilde{\Gamma}+\frac{\Gamma_{\downarrow}}{4}\upsilon$,
with $K\left(\omega_{d}\pm\omega\right)\approx K\left(\omega_{d}\right)\left(1\pm\epsilon+\upsilon\right)$.
Now, $\upsilon\propto\left(\frac{\omega}{\omega_{{\rm bath}}}\right)^{2}$,
i.e., the dephasing increases like $\Omega^{2}$. We have verified
numerically that the linewidths of the Mollow triplet taken with finite
$\upsilon$ increase linearly with $\Omega^{2}$, similar to experimental
results in comparable regimes \cite{newUlrich,EID}. Similarly to
the effect of $\epsilon$, this behavior exists also in the rotating
frame regime, and our proposed treatment allows one to dependably
obtain it in a broader parameter regime, namely weaker driving.

To conclude this Section, let us turn to an important regime of interest,
especially relevant in the context of single spin electron paramagnetic
resonance (EPR) experiments \cite{STMexperiment,Yang_TI_atoms,Willke2018},
where $\frac{1}{T_{1}}\ll\Omega\lesssim\frac{1}{T_{2}}$, i.e., very
small decay rates yet large dephasing coupling strength, which is
comparable to or exceeds the Rabi energy scale. The rotating frame
approach is ruled out immediately in such a regime. We have recently
demonstrated that experimental results for the aforementioned EPR
systems cannot be accounted for by the standard lab frame approach,
and that our generalized treatment enables capturing the important
observed features when taking into account the out-of-equilibrium
nature of the bath the system is coupled to, i.e., the electronic
reservoirs \cite{ShavitGH}. As an example for what is missed by the
lab frame approach, we consider $\Gamma_{\uparrow/\downarrow}=0$,
$\delta\omega=0$, and $\Gamma_{\pm}^{z}\equiv\Gamma_{0}^{z}\left(1\pm\Delta\frac{\omega}{\Gamma_{0}^{z}}\right)$,
i.e., we only consider the first order in $\omega$ correction to
the lab frame rates. In equilibrium, detailed balance dictates $\frac{\Delta}{\Gamma_{0}^{z}}\approx\frac{1}{T}$.
We calculate the spectral function \begin {widetext}
\begin{align}
g\left(\nu\right) & =\frac{\Gamma_{0}^{z}}{\left(2\Gamma_{0}^{z}\right)^{2}+\nu^{2}}\left[1-\left(\frac{\Delta\Omega}{\Gamma_{0}^{z}}\right)^{2}\right]+\nu^{2}\frac{\Gamma_{0}^{z}}{\left(2\Gamma_{0}^{z}\right)^{2}\nu^{2}+\left(\nu^{2}-\Omega^{2}\right)^{2}}+\nu\frac{\Delta\Omega}{\Gamma_{0}^{z}}\frac{\Omega\Gamma_{0}^{z}}{\left(2\Gamma_{0}^{z}\right)^{2}\nu^{2}+\left(\nu^{2}-\Omega^{2}\right)^{2}}.\label{eq:Doubletsky}
\end{align}
\end {widetext} For $\Delta=0$, we have the usual Mollow triplet
with peaks at $\nu=0,\pm\Omega$ for large driving amplitudes. As
$\Omega$ decreases, this latter term overshadows the first one, leading
to an apparent ``doublet'' structure, with a sharp dip at zero. The
third term in Eq. \eqref{eq:Doubletsky} is an asymmetry term, similar
to the one in Eq. \eqref{eq:gepsilon}. Fig. \ref{fig:InterestingRegimes}b
shows the shape of the spectrum with varying $\Omega$, far from the
rotating frame range of validity. It features clear asymmetry absent
from the lab frame approach.

\section {Applicabilty of the generalized quantum master equation}\label{Caveats}

As discussed above, our generalized master equations, Eqs.\eqref{eq:generalN}--\eqref{eq:generalA}
are not of the standard Lindblad form, unlike either lab frame or
rotating frame master equations. Keeping the important Markovian approximation,
as we still do in the generalized scheme, means that the master equation
now takes the Redfield form \cite{REDFIELD19651}, \textit{which does
not guarantee that the density matrix remains positive semi-definite}
\cite{MarkovianRelax,Memoryeffects}, though the Hermiticity and normalization
($\mathrm{Tr}\left\{ \rho\right\} =1$) are still preserved. Therefore,
when making use of the proposed generalized approach, one should understand
when it might become invalid, resulting in, e.g., ``negative probabilities''
(indicating unphysical results \cite{MarkovianRelax}). It should
be pointed out, however, that a master equation which has the Lindblad
form does not necessarily preclude non-physical behavior of the open
quantum system \cite{KosloffLevy_2014}. 

Since the trace of the density matrix is fixed at unity, as just stated,
for a two-level system its positivity is determined by its determinant:
A negative determinant would immediately signal that one of the density
matrix eigenvalue is negative (in addition to the second eigenvalue
being larger than $1$). The determinant can simply be written as
\begin{equation}
\det\rho=n\left(1-n\right)-\left|\alpha\right|^{2}.\label{eq:detDefine}
\end{equation}
We note that the possibility of $n<0$ or $n>1$ is also captured
by a negative density matrix determinant, as evident by \eqref{eq:detDefine}.
We find that although Eqs. \eqref{eq:labframeN}--\eqref{eq:labframeAlpha}
in principle allow these values of $n$, in practice one requires
unphysically large asymmetry parameters ($\epsilon,\Delta$) for this
to occur.

\begin{figure}
\begin{centering}
\includegraphics[scale=0.67]{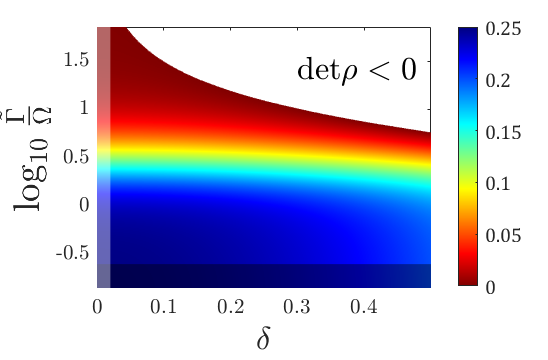}
\par\end{centering}
\caption{\label{fig:positivity} Determinant of the steady-state ($t\rightarrow\infty$)
density matrix in different regimes. The blank area is where the determinant
is negative, and therefore the solution is unphysical. The brighter
area to the left approximately corresponds to the rotating frame regime,
whereas the darker area at the bottom is properly
captured by the lab frame approach. Throughout this calculation we
used $\Gamma_{0}^{z}=2\Gamma_{\downarrow}$, $\Gamma_{\pm}^{z}=\Gamma_{0}^{z}\left(1\pm\delta\right)$,
and $\delta\omega=\Gamma_{\uparrow}=\epsilon_{e}=\epsilon=0$.}
\end{figure}

As an example, calculation of the determinant of $\rho$ in its steady-state
is presented in Fig. \ref{fig:positivity}. For illustrative purposes
the departure from the lab frame regime is achieved by forcing an
$\omega$-independent asymmetry in the longitudinal rates, $\Gamma_{\pm}^{z}=\Gamma_{0}^{z}\left(1\pm\delta\right)$,
whereas a small driving amplitude $\Omega$ compared to the decay
rates drives our master equation away from the rotating frame regime
of validity. Clearly, there exists a regime in parameter space where
positivity is not maintained (or is close to being lost), yet this
regime is far-removed from the scenarios that were previously accessible
under the approximations of the Lindbladian treatments. In other words,
in the area between the bottom of Fig. \ref{fig:positivity} and its
left hand side, representing the rotating and lab frame regimes respectively,
a significant amount of new ground is covered by the generalized approach,
allowing a better understanding of the open system dynamics in intermediate
regimes (as in Fig. \ref{fig:transientim}).

We note that restoring density matrix positivity in Redfield-type
master equations has been suggested to be possible \cite{SlippageRedfield}
by dividing the time evolution of the system into two parts: At initial
times the time evolution takes place with the full non-Markovian dynamics
(leading to a more complex master equation which is harder to handle).
After time $t_{b}$, which characterizes the bath relaxation time
and thus the scale at which any memory effects in the bath decay,
the unmodified Redfield time evolution is implemented. This can be
thought of as sort of a ``slippage'' of the initial conditions for
$\rho$. This sort of modification however, cannot resolve the discrepancy
shown in Fig. \ref{fig:positivity}, as it only pertains to the \textit{steady-state}
density matrix, which is independent of initial conditions. 

\begin{figure}
\begin{centering}
\includegraphics[viewport=38.3789bp 0bp 509bp 259bp,scale=0.55]{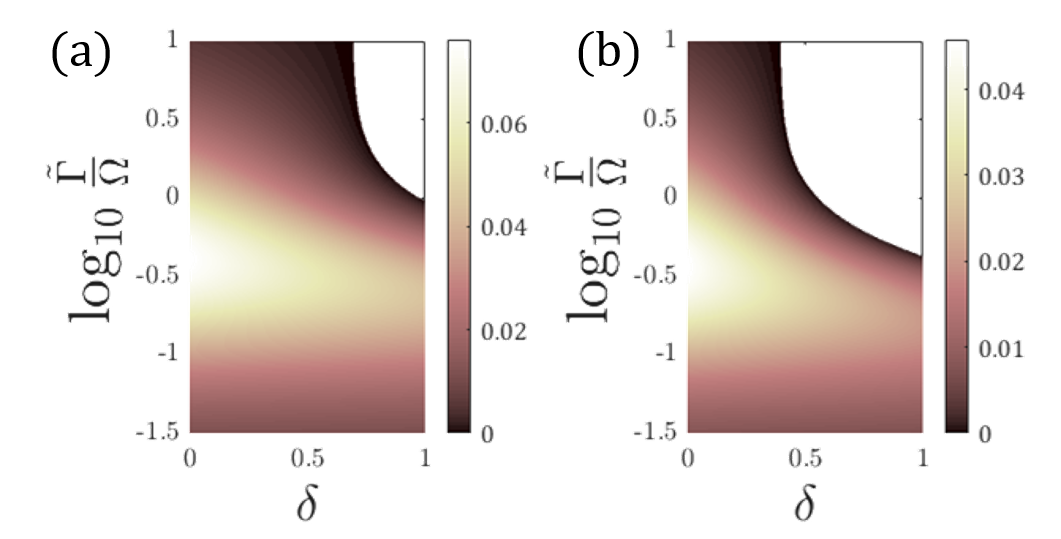}
\par\end{centering}
\caption{\label{fig:PowerFlow} Steady-state value of $\mathrm{Im}\left\{ \alpha\right\} $
as a function of driving strength and longitudinal asymmetry. The
blank areas represent $\mathrm{Im}\left\{ \alpha\right\} <0$, i.e.,
a negative power flow {[}Eq. \eqref{eq:power flow}{]}. Similarly
to Fig. \ref{fig:positivity}, the bottom of each panel corresponds
to the rotating frame regime, while the lab frame approach better
captures the leftmost $\delta=0$ parts. (a) $\Gamma_{\uparrow}=\frac{1}{4}\Gamma_{\downarrow}$;
(b) $\Gamma_{\uparrow}=\frac{1}{2}\Gamma_{\downarrow}$. Other parameters
are kept constant at $\Gamma_{0}^{z}=2\Gamma_{\downarrow}$, $\Gamma_{\pm}^{z}=\Gamma_{0}^{z}\left(1\pm\delta\right)$,
$\delta\omega=-\Gamma_{\downarrow}$, and $\epsilon_{e}=\epsilon=0$.}
\end{figure}

An additional criterion one may employ to determine the validity of
our proposed treatment is the rate at which energy introduced to the
system by the periodic drive is dissipated into the bath, which in
the steady-state is given by \cite{PowerFlow}
\begin{equation}
\mathcal{P}=\left\langle \frac{\partial}{\partial t}H_{D}\right\rangle =\hbar\omega_{d}\Omega\mathrm{Im}\left\{ \alpha\right\} ,\label{eq:power flow}
\end{equation}
where terms varying as $e^{\pm2i\omega_{d}t}$ were averaged to zero.
Except for the anomalous regime $\Gamma_{\uparrow}>\Gamma_{\downarrow}$
(where the bath causes net excitation of the system), this quantity
should remain non-negative to ensure that energy flows from the drive
through the qubit and into the bath. We find that for a combination
of ``red'' detuned driving ($\delta\omega<0$), finite $\frac{\Gamma_{\uparrow}}{\Gamma_{\downarrow}}$
ratio, and sufficient longitudinal asymmetry $\frac{\Gamma_{+}^{z}-\Gamma_{-}^{z}}{2\Gamma_{0}^{z}}$,
a regime appears where the generalized treatment results in apparent
negative power flow, see Fig. \ref{fig:PowerFlow}. However, once
again, the region in parameter space where the generalized treatment
breaks down is far from the reach of the previous approaches, and
a substantial formerly mistreated part of this parameter space can
now be more reliably accounted for.

\section {Conclusions}\label{Concluz}

A more general approach for treating a driven open quantum system
has been introduced and was shown to reduce to previous prevalent
Lindblad approaches in some specialized limits. The Lindbladian treatments
are limited in their applicability: One needs either the generalized
Rabi frequency $\omega$ to be sufficiently small such that spectral
properties of the bath do not change greatly under a shift of $\pm\omega$,
or that this frequency is much larger than all the dissipation rates.
The proposed treatment enables a seamless transition between these
two regimes, and allows a better understanding of their connection
to each other which was previously ambiguous, thus paving the way
towards the analysis of open quantum systems occupying new areas of
parameter space. In these previously unexplored parameter regimes
our proposed approach reveals, among other things, the intricate time-evolution
behavior of the crucial coherence terms, possibilities for population
inversion and EID even with intermediate driving amplitudes, and novel
structure in the resonance fluorescence spectrum. The generalized
approach we present in this work is free of the inapplicable approximations
which plague the existing approaches.

We demonstrate the relevance of our approach to experiment on quantum
dot qubits, where a crossover from the weak to the strong driving
regimes may occur, and the spectral content of the bath is not trivial
due to proximity to a cavity mode. We have shown that in such systems
the cavity-distorted bath spectral density may lead to an asymmetry
of the resonance fluorescence spectrum, even in the central peak,
an effect observed in experiments yet theoretically inaccessible in
either lab or rotating frame regimes.

Our proposed treatment is also needed for appropriately analyzing
single EPR experiments \cite{ShavitGH}, in addition to other systems
in the parameter regime $\frac{1}{T_{1}}\ll\Omega\lesssim\frac{1}{T_{2}}$.
In such regimes the Mollow triplet becomes a doublet, which becomes
visibly asymmetric for low enough temperatures, or in certain out-of-equilibrium
cases, as discussed in Ref. \cite{ShavitGH}.

Furthermore, as the recent trend of combining devices from different
disciplines and different regimes, is gradually accelerating \cite{HybridSystemsKlaus,Hybrid1,Hybrid2,Hybrid3},
the shift to a more unified perspective akin to the one presented
in this work may be unavoidable. Based on the principles shown in
this work, such an extension is straightforward, as our approach imposes
way less restrictions in terms of the relations between different
energy scales in the system.

\section* {Acknowledgments}

M. G. was supported by the Israel Science Foundation (Grant No. 227/15),
the German Israeli Foundation (Grant No. I-1259-303.10), the US-Israel
Binational Science Foundation (Grants No. 2014262 and 2016224), and
the Israel Ministry of Science and Technology (Contract No. 3-12419).
B. H. thanks stimulating discussions with A. Shnirman, G. Zar\'and,
and D. Cohen and acknowledges support by German-Israeli DIP project
(Hybrid devices: FO 703/2\textendash 1).

\appendix

\section {Deriving the generalized master equation}\label{deriving}

Starting from the driven Hamiltonian of a system coupled to a bath,
we present here the full derivation of the master equation\eqref{eq:generalN}--\eqref{eq:generalA},
and show that it generalizes the rotating frame approach. The full
Hamiltonian in the rotating frame with frequency $\omega_{d}$ is
\begin{align}
H & =\frac{1}{2}\delta\omega\sigma_{z}+\frac{1}{2}\Omega\sigma_{x}+H_{B}\\
 & -\frac{1}{2}\left(ae^{i\omega_{d}t}\sigma_{-}+a^{*}e^{-i\omega_{d}t}\sigma_{+}+a_{z}\sigma_{z}\right)\hat{B},
\end{align}
with $a=a_{x}+ia_{y}$. We can diagonalize the system part of this
Hamiltonian by transforming $H\rightarrow\tilde{H}=S^{-1}HS$, with
($\tan\beta\equiv\frac{\delta\omega}{\Omega}$)
\begin{equation}
S=\frac{1}{\sqrt{2}}\underbrace{\begin{pmatrix}\cos\frac{\beta}{2}+\sin\frac{\beta}{2} & -\cos\frac{\beta}{2}+\sin\frac{\beta}{2}\\
\cos\frac{\beta}{2}-\sin\frac{\beta}{2} & \cos\frac{\beta}{2}+\sin\frac{\beta}{2}
\end{pmatrix}}_{\mathrm{system\,subspace}}\otimes1_{\mathrm{bath\,subspace}}.\label{eq:SmatrixDiagonal}
\end{equation}
Thus, we have $\tilde{H}=\frac{1}{2}\omega\sigma_{z}-\left(A_{0}+A_{1}+A_{-1}\right)\hat{B}+H_{B},$
with\begin {subequations}

\begin{equation}
A_{0}=\left(\frac{\sin\beta}{2}a_{z}+\frac{\cos\beta}{4}\left(a^{*}e^{-i\omega_{d}t}+ae^{i\omega_{d}t}\right)\right)\sigma_{z},\label{eq:A0}
\end{equation}
\begin{equation}
A_{1}=\Lambda\left(t\right)\sigma_{-},
\end{equation}
\begin{equation}
A_{-1}=\Lambda^{*}\left(t\right)\sigma_{+},\label{eq:Am1}
\end{equation}
\end {subequations} where $\Lambda\left(t\right)=\frac{\sin\beta-1}{4}a^{*}e^{-i\omega_{d}t}+\frac{\sin\beta+1}{4}ae^{i\omega_{d}t}-\frac{\cos\beta}{2}a_{z}$.
Moving into the interaction picture of the system (made possible by
the diagonalization of the system Hamiltonian), $A_{\pm1}\left(t\right)$
pick up an additional $e^{\pm i\omega t}$ factor multiplying them.
In this picture the Markovian master equation reads \cite{REDFIELD19651}
\begin{equation}
\frac{d}{dt}\tilde{\rho}\left(t\right)=\sum_{j,k=-1}^{1}\int_{0}^{\infty}dsC\left(s\right)\left[A_{j}\left(t-s\right)\tilde{\rho}\left(t\right),A_{k}^{\dagger}\left(t\right)\right]+\mathrm{h.c.\,,}\label{eq:Redfield}
\end{equation}
with $C\left(\tau\right)\equiv\mathrm{Tr}_{B}\left\{ \rho_{B}\tilde{B}\left(t\right)\tilde{B}\left(t-\tau\right)\right\} $
the bath correlation function. We may now plug the $A_{j}\left(\tau\right)$
operators in the master equation, decompose the system density matrix
as $\tilde{\rho}=\frac{d+u}{2}+\frac{d-u}{2}\sigma_{z}+x\sigma_{-}+x^{*}\sigma_{+}$,
and perform the integral over time $s$ in \eqref{eq:Redfield}. Examination
of \eqref{eq:A0}--\eqref{eq:Am1} reveals that the only oscillation
frequencies which can appear in the master equation are
\[
0,\pm\omega,\pm2\omega,\pm\omega_{d},\pm\left(\omega_{d}\pm\omega\right),\pm2\omega_{d}.
\]
Keeping the RWA and the more general lab frame secular approximation
allows us to discard terms oscillating with frequencies that are in
the vicinity of $\omega_{d}$ or higher (we also assume $\omega_{d},\omega_{0}\gg\omega$),
leading to an equation of the form
\begin{align}
\frac{d}{dt}\tilde{\rho}\left(t\right) & =\mathcal{D}_{0}+\mathcal{D_{\omega}}+\mathcal{D}_{2\omega}+\mathrm{h.c.}\label{eq:rhoByDs}
\end{align}
Substituting the rates we have defined in the main text, these terms
read\begin {subequations} \begin {widetext}
\begin{eqnarray}
\mathcal{D}_{0} & = & -\left(d\sigma_{z}+x^{*}\sigma_{+}\right)\left(\frac{\left(\sin\beta-1\right)^{2}}{8}\Gamma_{\downarrow}\left(1-\epsilon\right)+\frac{\left(\sin\beta+1\right)^{2}}{8}\Gamma_{\uparrow}\left(1+\epsilon_{e}\right)+\frac{\cos^{2}\beta}{2}\Gamma_{-}^{z}\right)\nonumber \\
 &  & +\left(u\sigma_{z}-x\sigma_{-}\right)\left(\frac{\left(\sin\beta+1\right)^{2}}{8}\Gamma_{\downarrow}\left(1+\epsilon\right)+\frac{\left(\sin\beta-1\right)^{2}}{8}\Gamma_{\uparrow}\left(1-\epsilon_{e}\right)+\frac{\cos^{2}\beta}{2}\Gamma_{+}^{z}\right)\nonumber \\
 &  & -\left(x\sigma_{-}+x^{*}\sigma_{+}\right)\left(\sin^{2}\beta\Gamma_{0}^{z}+\frac{\cos^{2}\beta}{4}\left(\Gamma_{\downarrow}+\Gamma_{\uparrow}\right)\right),\label{eq:rotD0}
\end{eqnarray}
\begin{align}
{\cal D}_{\omega}= & -e^{i\omega t}d\sigma_{-}\cos\beta\left(\frac{\sin\beta-1}{4}\Gamma_{\downarrow}\left(1-\epsilon\right)+\frac{\sin\beta+1}{4}\Gamma_{\uparrow}\left(1+\epsilon_{e}\right)-\sin\beta\Gamma_{-}^{z}\right)\nonumber \\
 & +e^{-i\omega t}u\sigma_{+}\cos\beta\left(\frac{\sin\beta+1}{4}\Gamma_{\downarrow}\left(1+\epsilon\right)+\frac{\sin\beta-1}{4}\Gamma_{\uparrow}\left(1-\epsilon_{e}\right)-\sin\beta\Gamma_{+}^{z}\right)\nonumber \\
 & -e^{-i\omega t}\left(\sigma_{+}+x\sigma_{z}\right)\frac{\cos\beta}{2}\left(\frac{\sin\beta+1}{4}\Gamma_{\uparrow}+\frac{\sin\beta-1}{4}\Gamma_{\downarrow}-\sin\beta\Gamma_{0}^{z}\right)\nonumber \\
 & +e^{i\omega t}\left(\sigma_{-}-x^{*}\sigma_{z}\right)\frac{\cos\beta}{2}\left(\frac{\sin\beta-1}{4}\Gamma_{\uparrow}+\frac{\sin\beta+1}{4}\Gamma_{\downarrow}-\sin\beta\Gamma_{0}^{z}\right),\label{eq:rotDw}
\end{align}
\begin{eqnarray}
\mathcal{D}_{2\omega} & = & e^{2i\omega t}x^{*}\sigma_{-}\left(\frac{\sin^{2}\beta-1}{8}\Gamma_{\downarrow}\left(1-\epsilon\right)+\frac{\sin^{2}\beta-1}{8}\Gamma_{\uparrow}\left(1+\epsilon_{e}\right)+\frac{\cos^{2}\beta}{2}\Gamma_{-}^{z}\right)\nonumber \\
 &  & +e^{-2i\omega t}x\sigma_{+}\left(\frac{\sin^{2}\beta-1}{8}\Gamma_{\downarrow}\left(1+\epsilon\right)+\frac{\sin^{2}\beta-1}{8}\Gamma_{\uparrow}\left(1-\epsilon_{e}\right)+\frac{\cos^{2}\beta}{2}\Gamma_{+}^{z}\right),\label{eq:rotD2w}
\end{eqnarray}
\end {widetext}\end {subequations} where $\tilde{\rho}=\begin{pmatrix}d & x\\
x^{*} & u
\end{pmatrix}$. Note that the master equation for the rotating frame approach, Eqs.
\eqref{eq:rotMaster1}--\eqref{eq:rotMaster2}, are obtained by taking
the secular approximation with regards to $\omega$, i.e., $\mathcal{D}_{\omega},\mathcal{D}_{2\omega}\rightarrow0$,
and moving out of the interaction picture. However, by keeping all
these oscillating terms, we arrive at a more general form of $\frac{d}{dt}\tilde{\rho}$.
Following the unitary transformation defined in the main text as $e^{i\tilde{H}_{S}t}$,
we invert the diagonalization transformation $S$ {[}Eq. \eqref{eq:SmatrixDiagonal}{]},
in order to find the master equation in the original lab frame $\rho=\begin{pmatrix}1-n & \alpha\\
\alpha^{*} & n
\end{pmatrix}$, using \begin {subequations} 
\begin{equation}
n=\frac{1}{2}+\left(u-\frac{1}{2}\right)\sin\beta-\frac{x+x^{*}}{2}\cos\beta,
\end{equation}
\begin{equation}
\alpha=\left(u-\frac{1}{2}\right)\cos\beta+\frac{x+x^{*}}{2}\sin\beta-\frac{x-x^{*}}{2}.
\end{equation}
\end {subequations} This allows us to retrieve the full generalized
master equation in the main text, \eqref{eq:generalN}--\eqref{eq:generalA}.

For the sake of completeness, we bring here the master equations for
two limiting cases: (i) Pure dephasing only, i.e., $a=0$, which gives\begin {subequations}

\begin{eqnarray}
\frac{d}{dt}n & = & -i\Omega\frac{\alpha-\alpha^{*}}{2}\label{eq:pureDephaseN}
\end{eqnarray}
\begin{eqnarray}
\frac{d}{dt}\alpha & = & -\alpha\left[\left(\Gamma_{+}^{z}+\Gamma_{-}^{z}\right)\cos^{2}\beta+2\Gamma_{0}^{z}\sin^{2}\beta+i\delta\omega\right]\nonumber \\
 &  & +\left(n-\frac{1}{2}\right)\left[\frac{\sin2\beta}{2}\left(\Gamma_{+}^{z}+\Gamma_{-}^{z}-2\Gamma_{0}^{z}\right)-i\Omega\right]\nonumber \\
 &  & -\frac{\cos\beta}{2}\left(\Gamma_{+}^{z}-\Gamma_{-}^{z}\right),\label{eq:pureDephaseAlpha}
\end{eqnarray}
\end {subequations} and (ii) purely radiative decay ($a_{z}=0$),
giving 
\begin{eqnarray}
\frac{d}{dt}\alpha & = & -\alpha\left(\frac{\Gamma_{\downarrow}}{2}\left(1+\epsilon\sin\beta\right)+\frac{\Gamma_{\uparrow}}{2}\left(1+\epsilon_{e}\sin\beta\right)+i\delta\omega\right)\nonumber \\
 &  & -i\Omega\left(n-\frac{1}{2}\right)-\frac{\Gamma_{\downarrow}\epsilon-\Gamma_{\uparrow}\epsilon_{e}}{4}\cos\beta,\label{eq:raddecayAlpha}
\end{eqnarray}
with $\frac{d}{dt}n$ remaining the same as in Eq. \eqref{eq:generalN}.

\noindent \section {Correlation functions and the regression theorem}\label{triplet}

Consider a unitary time evolution $U\left(\tau,t\right)$ where $t$
is an initial time, at which the total density matrix is assumed to
be factorized $\rho_{E}\left(t\right)\otimes\rho_{S}\left(t\right)$,
into the environment (E) and system (S) density matrices, respectively.
This is in line with the derivation of a Lindblad type equations where
the onset of system-environment entanglement is at time $t=0$. Consider
the correlation of two system operators $B\left(t+\tau\right),A\left(t\right)$
(where the system has $N$ states)
\begin{multline}
C\left(\tau,t\right)=\langle B\left(t+\tau\right)A\left(t\right)\rangle\\
={\rm Tr}_{E,S}\left\{ U^{\dagger}\left(\tau,t\right)B\left(t\right)U\left(\tau,t\right)A\left(t\right)\rho_{E}\left(t\right)\rho_{S}\left(t\right)\right\} .
\end{multline}
We define $b\left(t\right)=A\left(t\right)\rho_{E}\left(t\right)\rho_{S}\left(t\right)$.
Each element of the $N\times N$ matrix $U\left(\tau,t\right)b\left(t\right)U^{\dagger}\left(\tau,t\right)$
(with implicit environment indices) is a linear combination of all
other elements. Hence, we can define a super-operator $K_{ij,lm}\left(\tau,t\right)$
that is an $N^{2}\times N^{2}$ matrix with system ``super-indices'',
such that 
\begin{equation}
\left[U\left(\tau,t\right)b\left(t\right)U^{\dagger}\left(\tau,t\right)\right]_{ij}=\sum_{lm}K_{ij,lm}\left(\tau,t\right)b_{lm}\left(t\right).\label{eq:r14}
\end{equation}
Since $A\left(t\right),B\left(t\right),\rho_{S}\left(t\right)$ are
independent of environment indices we can trace over the environment
and define a reduced time evolution for the system
\begin{equation}
K^{sys}\left(\tau,t\right)={\rm Tr}_{E}\left\{ K\left(\tau,t\right)\rho_{E}\left(t\right)\right\} .
\end{equation}
Taking now $b\left(t\right)=\rho_{E}\left(t\right)\otimes\rho_{S}\left(t\right)$
in Eq. \eqref{eq:r14} and tracing over $\rho_{E}$ shows that $K^{sys}\left(\tau,t\right)$
determines the time evolution of the system reduced density matrix
$\rho_{S}\left(t+\tau\right)=K^{sys}\left(\tau,t\right)\rho_{S}\left(t\right)$.
The correlation becomes 
\begin{equation}
C\left(\tau,t\right)={\rm Tr}_{S}\left\{ B\left(t\right)K^{sys}\left(\tau,t\right)A\left(t\right)\rho_{S}\left(t\right)\right\} 
\end{equation}
known as the quantum regression theorem \cite{QrEGRESSION}.

In particular, for two-level open quantum systems with $N=2$ we have
the formal form $K^{sys}\left(\tau\right)=e^{\hat{R}\tau}$, with
$\hat{R}$ being a 4x4 matrix determined by the master equation, e.g.,
Eqs. \eqref{eq:generalN}--\eqref{eq:generalA}. We calculate correlations
of the form $C_{-+}\left(\tau\right)=\langle\sigma_{-}\left(\tau\right)\sigma_{+}\left(0\right)\rangle$
for $\tau>0$ and $C_{-+}\left(-\tau\right)=C_{-+}^{*}\left(\tau\right)$
for $-\tau<0$, hence the Fourier transform
\begin{equation}
C_{-+}\left(\nu\right)=-2{\rm Re}\left\{ {\rm Tr}\left\{ \sigma_{-}\frac{1}{i\nu+\hat{R}}\sigma_{+}\rho_{\infty}\right\} \right\} ,
\end{equation}
where $\rho_{S}\left(t\right)\rightarrow\rho_{\infty}$ is usually
taken as the steady-state density matrix in a 4-vector form \cite{Rsupermatrix}.
We note that since equilibrium is achieved on a finite time scale
$\sim1/\Gamma$, on the long time scales of the Fourier transform
one can choose instead another initial $\rho$, as long as it is not
orthogonal to the steady-state one.

\bibliographystyle{auxiliary/apsrev4-1}
\phantomsection\addcontentsline{toc}{section}{\refname}\bibliography{auxiliary/references}

\end{document}